**Numerical Investigations of Kuiper Belt Binaries.** R. C. Nazzario and T. W. Hyde

Center for Astrophysics, Space Physics and Engineering Research, Baylor University, Waco, TX 76798-7310, USA, Truell_Hyde@Baylor.edu/ phone: 254-710-3763

**Introduction:** Observations of the Kuiper Belt indicate that a larger than expected percentage of KBO's (approximately 8 out of 500) are in binary pairs. The formation and survival of such objects presents a conundrum [1]. Two competing theories have been postulated to try to solve this problem. One entails the physical collision of bodies [2] while the other utilizes dynamical friction or a third body to dissipate excess momentum and energy from the system [3]. Although in general known binaries tend to differ significantly in mass, such as seen in the Earth-Moon or asteroid binary systems [4], Kuiper binaries discovered to date tend to instead be of similar size [5, 6]. This paper investigates the stability, development and lifetimes for Kuiper Belt binaries by tracking their orbital dynamics and subsequent evolution. Section two details the numerical model while Section three discusses the initial conditions. Finally, in Section four the results are discussed with Section five containing the conclusions.

**Numerical Model:** The numerical method employed in this work is a 5$^{th}$ order Runga-Kutta algorithm [7] based on the Butcher's scheme [8]. A fixed time step of two days was used in order to reduce truncation and round-off errors while still yielding reasonable CPU run times over a simulation period of 1000 years. The forces considered include the gravitational attraction of the Sun, the seven major planets (Venus through Neptune) and all other KBO's with the corresponding accelerations due to these forces given by Eq. (1).

$$\vec{a}_i = -\frac{GM_{Sun}}{r^2}\hat{r} - \sum_{i=1}^{n}\sum_{\substack{j=1 \\ i \neq j}}^{n}\frac{Gm_j}{r_{ij}^2}\hat{r}_{ij} \qquad (1)$$

Particles are considered removed from the system when they venture beyond 100 AU or enter the sphere of influence (SOI) of one of the seven major planets. The SOI is defined as the distance from the secondary attractor, where bodies will interact with one another as opposed to being primarily acted upon by the Sun. This is given by Eq. (2),

$$r_{SOI} \approx \left(\frac{M_{Secondary}}{M_{Primary}}\right)^{2/5} r \qquad (2)$$

where $M_{Secondary}$ is the mass of the planet, $M_{Primary}$ is the mass of the Sun, and r is the distance of separation between the primary and secondary bodies. Collisions between objects are allowed as is any subsequent KBO size change.

**Initial Conditions:** The Kuiper Belt is divided into three distinct populations: the classic Kuiper Belt Objects, the resonant KBO's, and the scattered KBO's. These simulations are representative of the classic Kuiper Belt population (containing 50% of all KBO's) where KBO orbits lie in the ecliptic plane and eccentricities are generally less than 0.2 [4]. This specific KBO population has a size distribution for which the number density of 100 km particles is a maximum [5]. Therefore, a radius of 100 km was chosen as the radius for all simulated KBO's. A mean density of 3.5 g/cm$^3$ was assumed (corresponding roughly to dirty water ice) yielding individual KBO masses of 1.47x10$^{19}$ kg.

A total of 4000 KBO's were initially arranged as 2000 binary pairs randomly distributed between 30 and 35 AU. These 2000 binary pairs were divided into 10 sets of 200 and assigned orbital eccentricities between 0.0 to 0.9 in steps of 0.1. Each pair had their initial position (and hence velocity) randomized about their center of mass with the center of mass then given the velocity it would have had were it in a circular orbit about the Sun. Figure 1 shows this initial particle distribution.

**Results:** Once established, binaries were found to be remarkably stable and were only rarely disrupted during their subsequent evolution. Of the 2000 pairs simulated, 140 collided with Neptune during the 1000-year simulation period with no collisions occurring between KBO's. Four quaternary systems and two tertiary systems were formed with 150 years left in the simulation, remaining stable throughout the remainder of the 1000-year time period. It appears the initial eccentricity of the KBO binary may be a factor in determining whether such higher-level systems will form since initial eccentricities below 0.4 resulted in five of the six multi-particle systems discovered. It is also interesting to note that throughout the 1000-year simulation, zero KBO's were ejected into the Oort cloud.

The final positions for all KBO's tracked in this simulation are shown in Fig. 2 with their average eccentricity and their minimum and maximum eccentricities given in Table 1. As can be seen, KBO binaries having smaller initial eccentricities generally suffer larger perturbations than do those with initially larger eccentricities.

**Conclusions:** Presently, there are only 8 known KBO binaries in the Kuiper Belt constituting approximately 4% of the whole. These simulations predict binary disruption rates of approximately 7% per 1000 years, implying KBO binary lifetimes in agreement with previous results [9, 10]. Additionally, the appearance of coupled systems consisting of more than two objects agrees with previous studies that postulated systems of higher multiplicity could occur [3]. It has also recently been shown that the particle distribution of KBO's over a specific spatial region is related to transient binary formation [11]. KBO densities on the order of $12/AU^2$ resulted in transient binaries being formed with negligible binary formation found for KBO densities lower than approximately $2/AU^2$ (corresponding to the 0.35% found in this research). Thus, the formation of tertiary and quaternary systems offers a possible mechanism for providing the initial density of KBO's required for the formation of binary systems. The above is also in agreement with current predictions that the Kuiper Belt was originally up to 100 times more massive than it is presently. These possibilities will be discussed in upcoming publications.

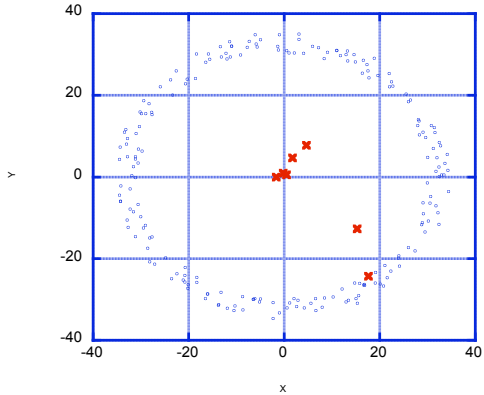

Figure 1. Initial KBO positions. X's represent the major planets while dots represent the KBO's.

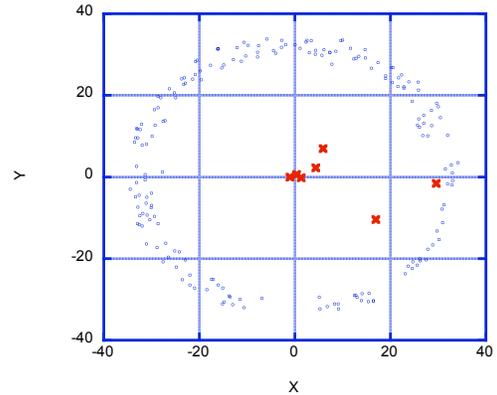

Figure 2. Final KBO positions. X's represent the major planets while dots represent the KBO's

Table 1. Final average eccentricities compared with the initial, minimum and maximum eccentricity for each set.

| Initial Eccentricity | Final Eccentricity | Minimum Final Eccentricity | Maximum Final Eccentricity |
|---|---|---|---|
| 0.0 | 0.008 | 0.000 | 0.763 |
| 0.1 | 0.106 | 0.090 | 0.607 |
| 0.2 | 0.202 | 0.177 | 0.222 |
| 0.3 | 0.306 | 0.267 | 0.890 |
| 0.4 | 0.405 | 0.360 | 0.586 |
| 0.5 | 0.501 | 0.501 | 0.547 |
| 0.6 | 0.604 | 0.554 | 0.648 |
| 0.7 | 0.703 | 0.621 | 0.749 |
| 0.8 | 0.802 | 0.714 | 8.450 |
| 0.9 | 0.901 | 0.800 | 0.936 |